\documentclass{www2012-accepted}

\usepackage{graphicx}
\usepackage[usenames,dvipsnames]{color}
\usepackage{booktabs}
\usepackage{amsfonts}
\usepackage{amsmath}
\usepackage[caption=false]{subfig}

\definecolor{darkgreen}{rgb}{0.15,0.4,0.15}

\begin{document}

\clubpenalty=10000 
\widowpenalty = 10000

\title{The Role of Social Networks in Information Diffusion}

\numberofauthors{4}
\author{
 \alignauthor Eytan Bakshy\thanks{Part of this research was performed while the author was a student at the University of Michigan.}\\
    \affaddr{Facebook}\\
    \affaddr{1601 Willow Rd.}\\
    \affaddr{Menlo Park, CA 94025}\\
    \email{ebakshy@fb.com}
 \alignauthor Itamar Rosenn\\
    \affaddr{Facebook}\\
    \affaddr{1601 Willow Rd.}\\
    \affaddr{Menlo Park, CA 94025}\\
    \email{itamar@fb.com}
 \and
 \alignauthor Cameron Marlow\\
    \affaddr{Facebook}\\
    \affaddr{1601 Willow Rd.}\\
    \affaddr{Menlo Park, CA 94025}\\
    \email{cameron@fb.com}
 \alignauthor Lada Adamic\\
    \affaddr{University of Michigan}\\
    \affaddr{105 S. State St.}\\
    \affaddr{Ann Arbor, MI 48104}\\
    \email{ladamic@umich.edu}
}

\maketitle
\begin{abstract}
Online social networking technologies enable individuals to simultaneously share information with any number of peers. Quantifying the causal effect of these mediums on the dissemination of information requires not only identification of who influences whom, but also of whether individuals would still propagate information in the absence of social signals about that information.  We examine the role of social networks in online information diffusion with a large-scale field experiment that randomizes exposure to signals about friends' information sharing among 253 million subjects in situ. Those who are exposed are significantly more likely to spread information, and do so sooner than those who are not exposed. We further examine the relative role of strong and weak ties in information propagation. We show that, although stronger ties are individually more influential, it is the more abundant weak ties who are responsible for the propagation of novel information. This suggests that weak ties may play a more dominant role in the dissemination of information online than currently believed.
\end{abstract}



\section{Introduction}
Social influence can play a crucial role in a range of behavioral
phenomena, from the dissemination of information, to the adoption of
political opinions and technologies~\cite{granovetter1978,watts98collective}, which are increasingly mediated through online systems~\cite{fox2011,purcell2010}.
Despite the wide availability of data from online social networks, identifying influence remains a challenge.
Individuals tend to engage in similar activities as their peers, so it is often impossible to determine from observational data whether a correlation between two individuals' behaviors exists because they are similar or because one person's behavior has influenced the other~\cite{aral2009,manski1993,shalizi2011}.
In the context of information diffusion, two people may disseminate the same information as each other because they possess the same information sources, such as web sites or television, that they consume regularly~\cite{adar2009,purcell2010}.

Moreover, homophily -- the tendency of individuals with similar characteristics to associate with one another~\cite{adamic01friends,kossinets09homophily,mcpherson2001} -- creates difficulties for measuring the relative role of strong and weak ties in information diffusion, since people are more similar to those with whom they interact often~\cite{granovetter1973,mcpherson2001}.
On one hand, pairs of individuals who interact more often have greater opportunity to influence one another and have more aligned interests, increasing the chances of contagion~\cite{brown1987,hill2006}.
However, this commonality amplifies the potential for confounds: those who interact more often are more likely to have increasingly similar information sources.  
As a result, inferences made from observational data may overstate the importance of strong ties in information spread. 
Conversely, individuals who interact infrequently have more diverse social networks that provide access to novel information~\cite{burt1992,granovetter1973}.
But because contact between such ties is intermittent, and the individuals tend to be dissimilar, any particular piece of information is less likely to flow across weak ties~\cite{centola2007,onnela2007}.  
Historical attempts to collect data on how often pairs of individuals communicate and where they get their information have been prone to biases~\cite{bernard1984,marin1981}, further obscuring the empirical relationship between tie strength and diffusion.

Confounding factors related to homophily can be addressed using controlled experiments, but experimental work has thus far been confined to the spread of highly specific information within limited populations~\cite{aral2011creating,centola2010}.  
In order to understand how information spreads in a real-world environment, we wish to examine a setting where a large population of individuals frequently exchange information with their peers.
Facebook is the most widely used social networking service in the world, with over 800 million people using the service each month. 
For example, in the United States, 54\% of adult Internet users are on Facebook~\cite{hampton2011}.
Those American users on average maintain 48\% of their real world contacts on the site~\cite{hampton2011}, and many of these individuals regularly exchange news items with their contacts~\cite{purcell2010}. 
In addition, interaction among users is well correlated with self-reported intimacy~\cite{gilbert2009}. 
Thus, Facebook represents a broad online population of individuals whose online personal networks reflect their real-world connections, making it an ideal environment to study information contagion.

We use an experimental approach on Facebook to measure the spread of information sharing behaviors.
The experiment randomizes whether individuals are exposed via Facebook to information about their friends' sharing behavior, thereby devising two worlds under which information spreads: one in which certain information can only be acquired external to Facebook, and another in which information can be acquired within or external to Facebook. By comparing the behavior of individuals within these two conditions, we can determine the causal effect of the medium on information sharing.

The remainder of this paper is organized as follows.  We further motivate our study with additional related work in Section~\ref{sec:related_work}.  Our experimental design is described in Section~\ref{sec:experimental_design}.  Then, in Section~\ref{sec:exposure} we discuss the causal effect of exposure to content on the newsfeed, and how friends' sharing behavior is correlated in time, irrespective of social influence via the newsfeed.  Furthermore, we show that multiple sharing friends are predictive of sharing behavior regardless of exposure on the feed, and that additional friends do indeed have an increasing causal effect on the propensity to share. In Section~\ref{sec:tie_strength} we discuss how tie strength relates to influence and information diffusion.  We show that users are more likely to have the same information sources as their close friends, and that simultaneously, these close friends are more likely to influence subjects.  Using the empirical distribution of tie strength in the network, we go on to compute the overall effect of strong and weak ties on the spread of information in the network.  Finally, we discuss the implications of our work in Section~\ref{sec:discussion}.

\section{Related Work}\label{sec:related_work}
Online networks are focused on sharing information, and as such, have been studied extensively in the context of information diffusion. Diffusion and influence have been modeled in blogs~\cite{adar_05,Gomez2010,gruhl2004information}, email~\cite{liben2008tracing}, and sites such as Twitter, Digg, and Flickr~\cite{bakshy2011wsdm,Goyal:2010:LIP:1718487.1718518,lerman2010information}.  One particularly salient characteristic of diffusion behavior is the correlation between the number of friends engaging in a behavior and the probability of adopting the behavior.  This relationship has been observed in many online contexts, from the joining of LiveJournal groups~\cite{backstrom06}, to the bookmarking of photos~\cite{cha2009}, and the adoption of user-created content~\cite{bakshy2009}.  However, as Anagnostopoulos, et al.~\cite{Anagnostopoulos08kdd} point out, individuals may be more likely to exhibit the same behavior as their friends because of homophily rather than as a result of peer influence. Statistical techniques such as permutation tests and matched sampling~\cite{aral2009} help control for confounds, but ultimately cannot resolve this fundamental problem~\cite{shalizi2011}. 

Not all diffusion studies must infer whether one individual influenced
another.  For example, Leskovec et al.~\cite{leskovec06viral} study
the explicit graph of product recommendations, Sun et
al.~\cite{sun_09} study cascading in page fanning, and Bakshy et
al.~\cite{bakshy2009} examine the exchange of user-created content.
However, in all these studies, even if the source of a particular
contagion event is a friend, such data does not tell us about the
relative importance of social networks in information diffusion.   For
example, consider the spread of news.  In Bradley Greenberg's classsic
study of media contagion~\cite{greenberg1964}, 50\% of respondents
learned about the Kennedy assassination via interpersonal
ties. Despite the substantial word-of-mouth spread, it is clear that all
of the respondents would have gotten
the news at a slightly later point in time (perhaps from the very same media
outlets as their contacts), had they not communicated with their
peers. Therefore, a complete understanding of the importance of social networks in information diffusion not only requires us to identify sources of interpersonal contagion, but also requires a counterfactual understanding of what would happen if certain interactions did not take place.

\begin{figure}[h]
\begin{center}
\centerline{\includegraphics[width=0.48\textwidth]{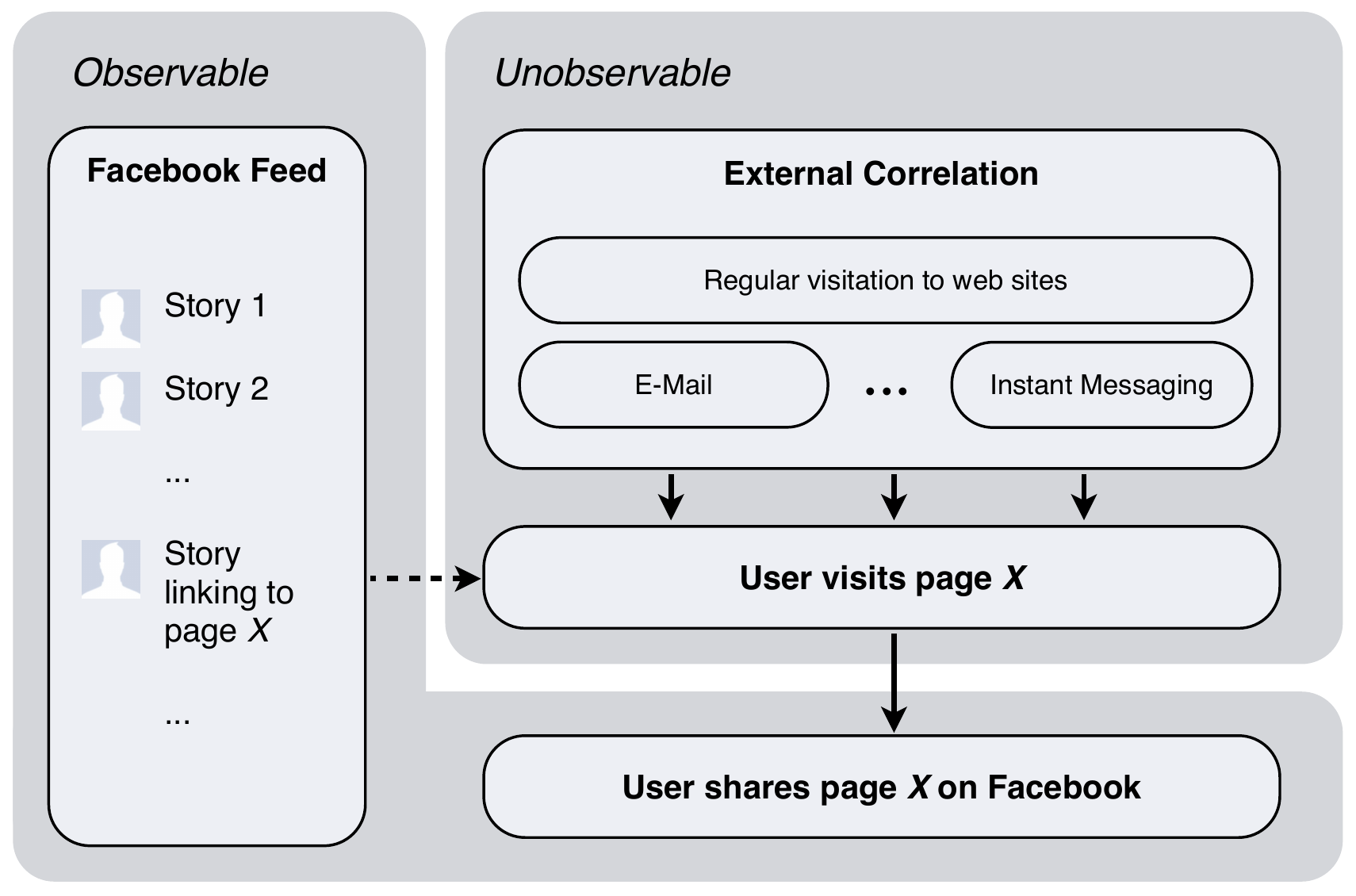}}  
\caption{Causal relationships that explain diffusion-like phenomena. Information presented in users' news feeds  and other sharing behavior on \texttt{facebook.com} are observed.  External events that cause users to be exposed to information outside of Facebook cannot be observed and may explain their sharing behavior.  Our experiment blocks the causal relationship (dashed arrow) between the Facebook newsfeed and user visitation by randomly removing stories about friends' sharing behavior in subjects' feeds.  Thus, our experiment allows us to compare situations where both influence via the feed and external correlations exist (the {\it feed} condition), to situations in which only external correlations exist (the {\it no feed} condition).}
\label{fig:causal}
\end{center}
\end{figure}

\begin{figure*}[ht]
\begin{center}
\centering
\subfloat[]{\includegraphics[width=0.47\textwidth]{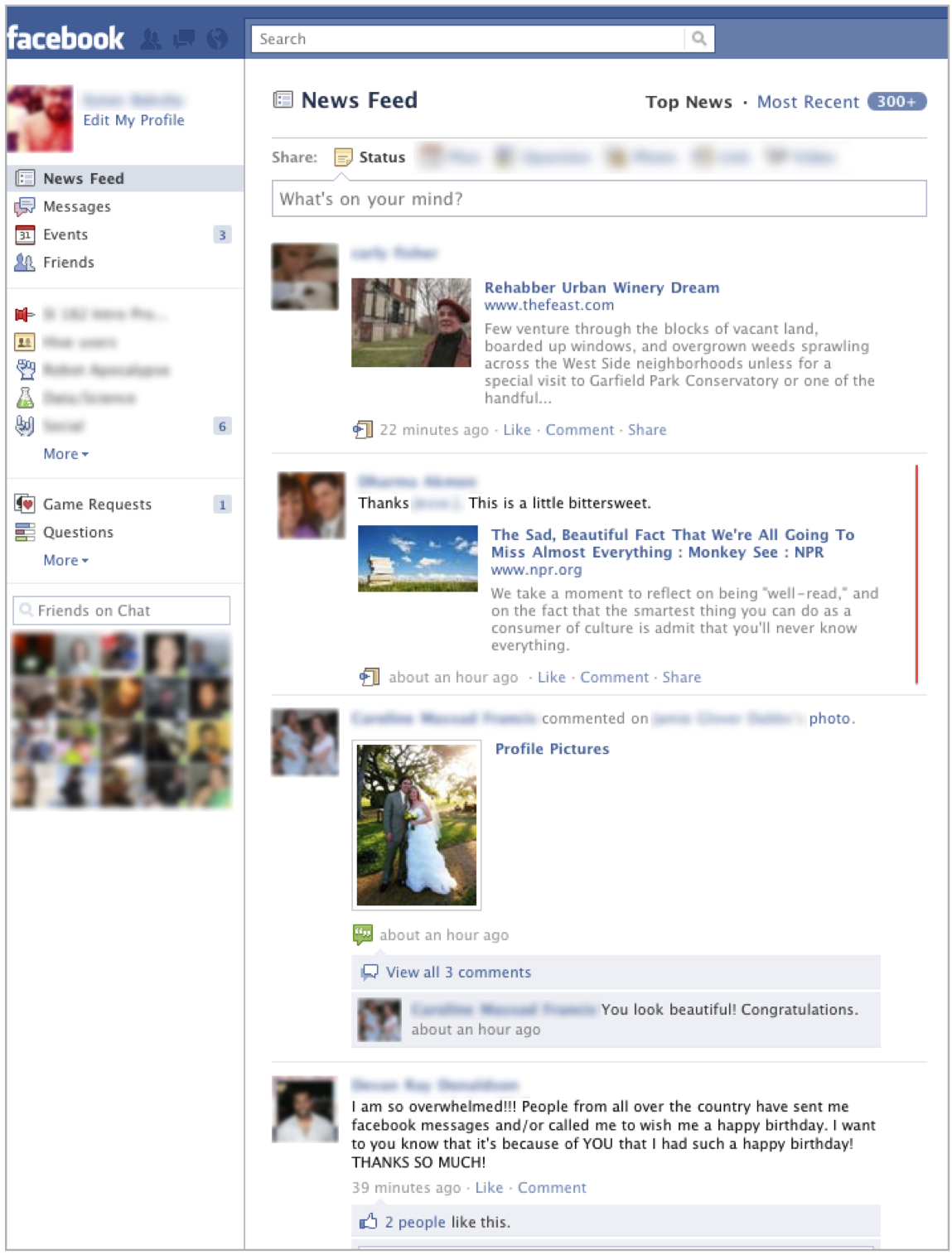}}
\hspace{24pt}
\subfloat[]{\includegraphics[width=0.47\textwidth]{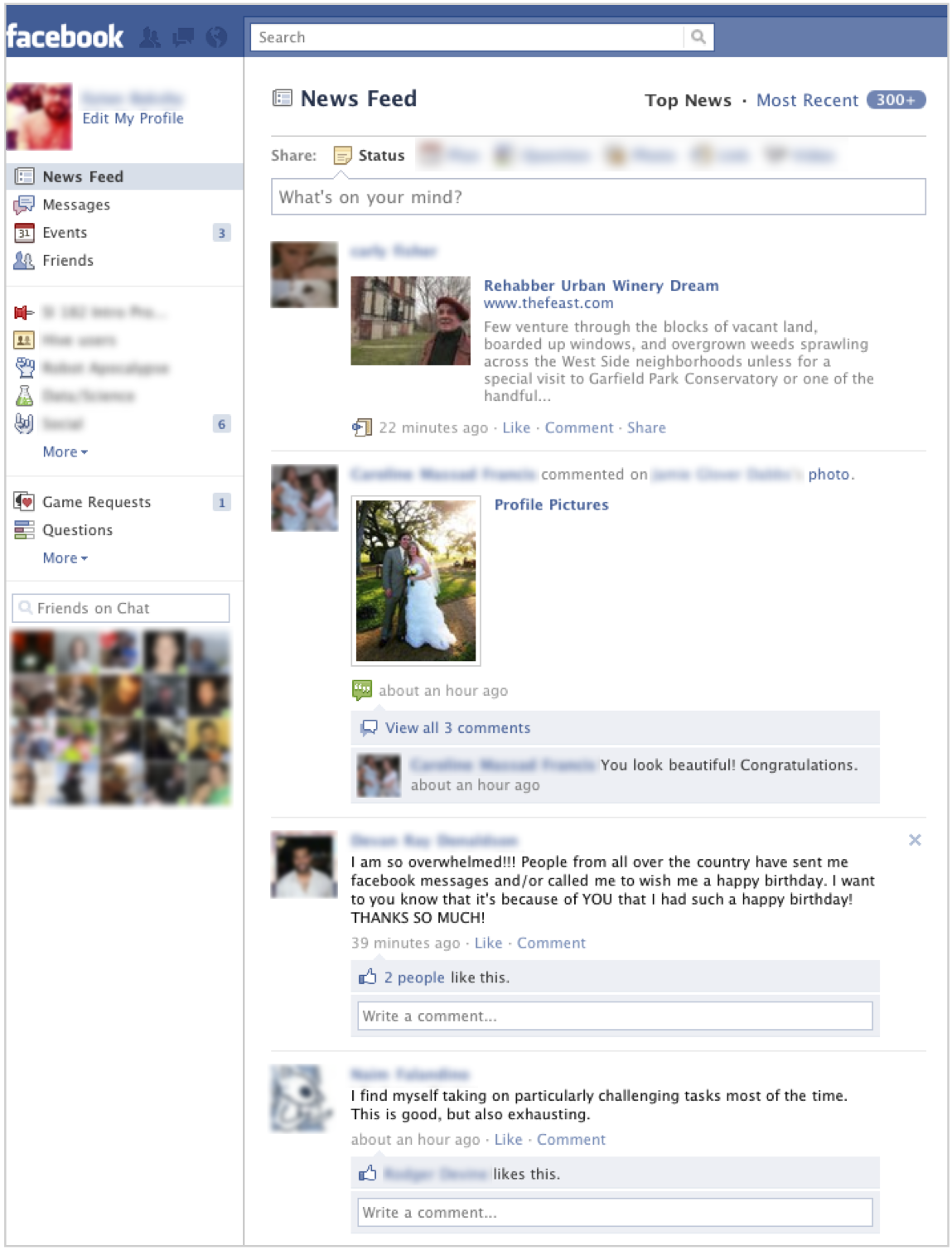}}
\caption{An example of the Facebook News Feed interface for a hypothetical subject who has a link (highlighted in red) assigned to the (a) {\it feed} or (b) {\it no feed} condition.}
\label{fig:screenshot_feed}
\end{center}
\end{figure*}

\section{Experimental Design and Data}\label{sec:experimental_design}
Facebook users primarily interact with information through an aggregated history of their friends' recent activity (stories), called the News Feed, or simply feed for short.
Some of these stories contain links to content on the Web, uniquely identified by URLs.  Our experiment evaluates how much exposure to a URL on the feed increases an individual's propensity to share that URL, beyond correlations that one might expect among Facebook friends. For example, friends with whom a user interacts more often may be more likely to visit sites that the user also visits. As a result, those friends may be more likely to share the same URL as the user before she has the opportunity to share that content herself. Additional unobserved correlations may arise due to external influence via e-mail, instant messaging, and other social networking sites.  These causal relationships are illustrated in Figure~\ref{fig:causal}.  From the figure, one can see that all unobservable correlations can be identified by blocking the causal relationship between the Facebook feed and sharing.  Our experiment therefore randomizes subjects with respect to whether they receive social signals about friends' sharing behavior of certain Web pages via the Facebook feed.

\subsection{Assignment Procedure}
Subject-URL pairs are randomly assigned at the time of display to either the {\it no feed} or the {\it feed} condition.  Stories that contain links to a URL assigned to the {\it no feed} condition for the subject are never displayed in the subject's feed.  Those assigned to the {\it feed} condition are not removed from the feed, and appear in the subject's feed as normal (Figure~\ref{fig:screenshot_feed}).  Pairs are deterministically assigned to a condition at the time of display, so any subsequent share of the same URL by any of a subject's friends is also assigned to the same condition.  To improve the statistical power of our results, twice as many pairs were assigned to the {\it no feed} condition. Because removal from the feed occurs on a subject-URL basis, and we include only a small fraction of subject-URL pairs in the {\it no feed} condition, a shared URL is on average delivered to over 99\% of its potential targets.

All activity relating to subject-URL pairs assigned to either experimental condition is logged, including feed exposures, censored exposures, and clicks to the URL (from the feed or other sources, like messaging).  Directed shares, such as a link that is included in a private Facebook message or explicitly posted on a friend's wall, are not affected by the assignment procedure.  If a subject-URL pair is assigned to an experimental condition, and the subject clicks on content containing that URL in any interface other than the feed, that subject-URL pair is removed from the experiment.  Our experiment, which took place over the span of seven weeks, includes 253,238,367 subjects, 75,888,466 URLs, and 1,168,633,941 unique subject-URL pairs.

\subsection{Ensuring Data Quality}
Threats to data quality include using content that was or may have
been previously seen by subjects on Facebook prior to the experiment,
content that subjects may have seen through interfaces on Facebook
other than feed, spam, and malicious content.  We address these issues
in a number of ways.  First, we only consider content that was shared
by the subjects' friends only after the start of the experiment. This
enables our experiment to accurately capture the first time a subject
is exposed to a link in the feed, and ensures that URLs in our
experiment more accurately reflect content that is primarily being
shared contemporaneously with the timing of the experiment. We also
exclude potential subject-URL pairs where the subject had previously
clicked on the URL via any interface on the site at any time up to two
months prior to exposure, or any interface other than the feed for
content assigned to the {\it no feed} condition.  Finally, we use the
Facebook's site integrity system~\cite{stein2011} to classify and remove URLs that may not reflect ordinary users' purposeful intentions of distributing content to their friends.

\subsection{Population}
The experimental population consists of a random sample of all Facebook users who visited the site between August 14th to October 4th 2010, and had at least one friend sharing a link.  At the time of the experiment, there were approximately 500 million Facebook users logging in at least once a month.  Our sample consists of approximately 253 million of these users.  All Facebook users report their age and gender, and a user's country of residence can be inferred from the IP address with which she accesses the site.   In our sample, the median and average age of subjects is 26 and 29.3, respectively.  Subjects originate from 236 countries and territories, 44 of which have one million or more subjects. Additional summary statistics are given in Table~\ref{tab:demographics}, and show that subjects are assigned to the conditions in a balanced fashion. 

\begin{table}[tp]
\begin{center}
\footnotesize
\begin{tabular}{lll}
Demographic Feature & {\bf feed} & {\bf no feed} \\
(\% of subjects) &&\\
\toprule
Gender &  \\
~{\sc Female} & 51.6\% & 51.4\% \\
~{\sc Male} & 46.7\% & 47.0\% \\
~{\sc Unspecified} & 1.5\% & 1.5\% \\
Age &  \\
~{\sc 17 or younger} & 12.8\% & 13.1\% \\
~{\sc 18-25} & 36.4\% & 36.1\% \\
~{\sc 26-35} & 27.2\% & 26.9\% \\
~{\sc 36-45} & 13.0\% & 12.9\% \\
~{\sc 46 or older} & 10.6\% & 10.9\% \\
Country (top 10 \& other) & \\
~{\sc United States} & 28.9\% & 29.1\% \\ 
~{\sc Turkey} & 6.1\% & 5.8\% \\ 
~{\sc Great Britain} & 5.1\% & 5.2\% \\ 
~{\sc Italy} & 4.2\% & 4.1\% \\ 
~{\sc France} & 3.8\% & 3.9\% \\ 
~{\sc Canada} & 3.7\% & 3.8\% \\ 
~{\sc Indonesia} & 3.7\% & 3.5\% \\ 
~{\sc Philippines} & 2.1\% & 2.3\% \\ 
~{\sc Germany} & 2.3\% & 2.3\% \\ 
~{\sc Mexico} & 2.0\% & 2.1\% \\
~{\sc 226 Others} & 37.5\% & 37.7\%
\end{tabular}
\caption {Summary of demographic features of subjects assigned to the {\it feed} ($N=160,688,092$) and {\it no feed} ($N=218,743,932$) condition. Some subjects may appear in both columns.
}
\label{tab:demographics}
\end{center}
  
\end{table}

\subsection{Evaluating Outcomes}
The assignment procedure allows us to directly compare the overall
probability that subjects share links they were or were not exposed to
on the feed.  The causal effect of exposure via the Facebook feed on
sharing is simply the expected probability of sharing in the {\it
  feed} condition minus the expected probability in the {\it no feed}
condition.  This quantity, known as the average treatment effect on
the treated (or alternatively, the absolute risk increase), can vary when conditioning on other variables, including the number of friends and tie strength, which are analyzed in Sections~\ref{sec:exposure} and~\ref{sec:tie_strength}.  Alternatively, the difference in probabilities can be viewed as a ratio (the relative risk ratio), which quantifies how many times more likely an individual is to share as a result of being exposed to content on the feed.

Although the assignment is completely random, subjects and URLs may differ in ways that impact our measurements.  For example, certain users may be highly active on Facebook, so that they are assigned to experimental conditions more often than other users.  If these users were to vary significantly in terms of their information sharing propensities, such as sharing or re-sharing greater or fewer links than others, the disproportionate inclusion of these users may bias our measurements and threaten the population validity of our findings.  Similarly, very popular URLs may also introduce biases; they may be more or less likely to be re-shared because of their inherent appeal or more likely to be discovered independently of Facebook because of their relative popularity amongst friends.

To provide control for these biases, we use bootstrapped averages clustered by the subject or URL.  We find that in all of our analyses, clustering by the URL rather than the subject yields nearly identical probability estimates that have marginally wider confidence intervals, so we have chosen to present our results using means and 95\% confidence intervals clustered by URL.  Risk ratios are obtained using the 95\% bootstrapped confidence intervals of likelihood of sharing in the {\it feed} and {\it no feed} conditions.  To compute the lower bound of the ratio, we divide the lower bound of the probability of sharing in the {\it feed} condition by the upper bound for the {\it no feed} condition.  The upper bound of the ratio is computed by dividing the upper bound in the {\it feed} condition by the lower bound of the {\it no feed} condition.  The additive analog of the same procedure is used to obtain confidence intervals for probability differences.

\section{How Exposure to Social Signals Affects Diffusion}\label{sec:exposure}
We find that subjects who are exposed to signals about friends' sharing behavior are several times more likely to share that same information, and share sooner than those who are not exposed.
To measure the relative increase in sharing due to exposure, we compute the risk ratio:  
the likelihood of sharing in the {\it feed} condition (0.191\%) divided by the likelihood of sharing in the {\it no feed condition} (0.025\%), 
and find that individuals in the {\it feed} condition are 7.37 times more likely share ($95\%\ CI = [7.23, 7.72]$).
Although the probability of sharing upon exposure may appear small, it is important to note that individuals have hundreds of contacts online who may see their link, and that on average one out of every 12.5 URLs that are clicked on in the {\it feed} condition are subsequently re-shared.

\begin{figure}[!h]
\begin{center}
\begin{tabular}{cc}
  \includegraphics[width=0.8\columnwidth]{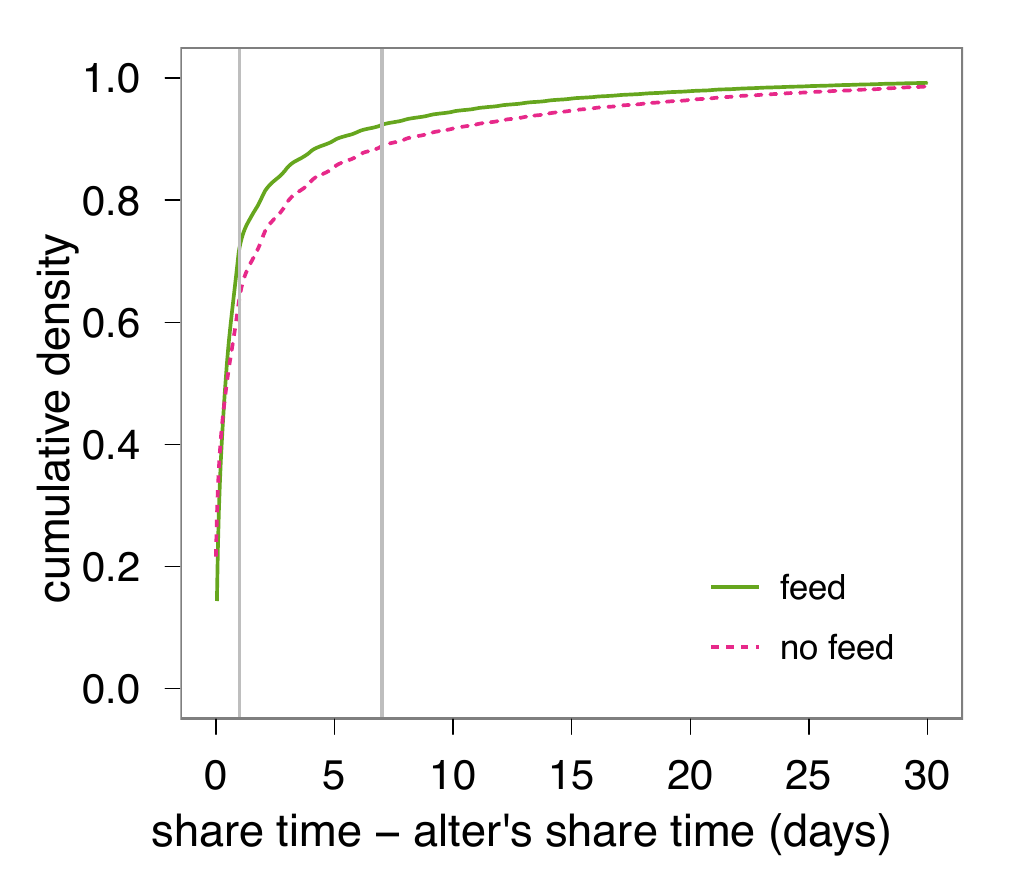} \\
(a) \\
  \includegraphics[width=0.8\columnwidth]{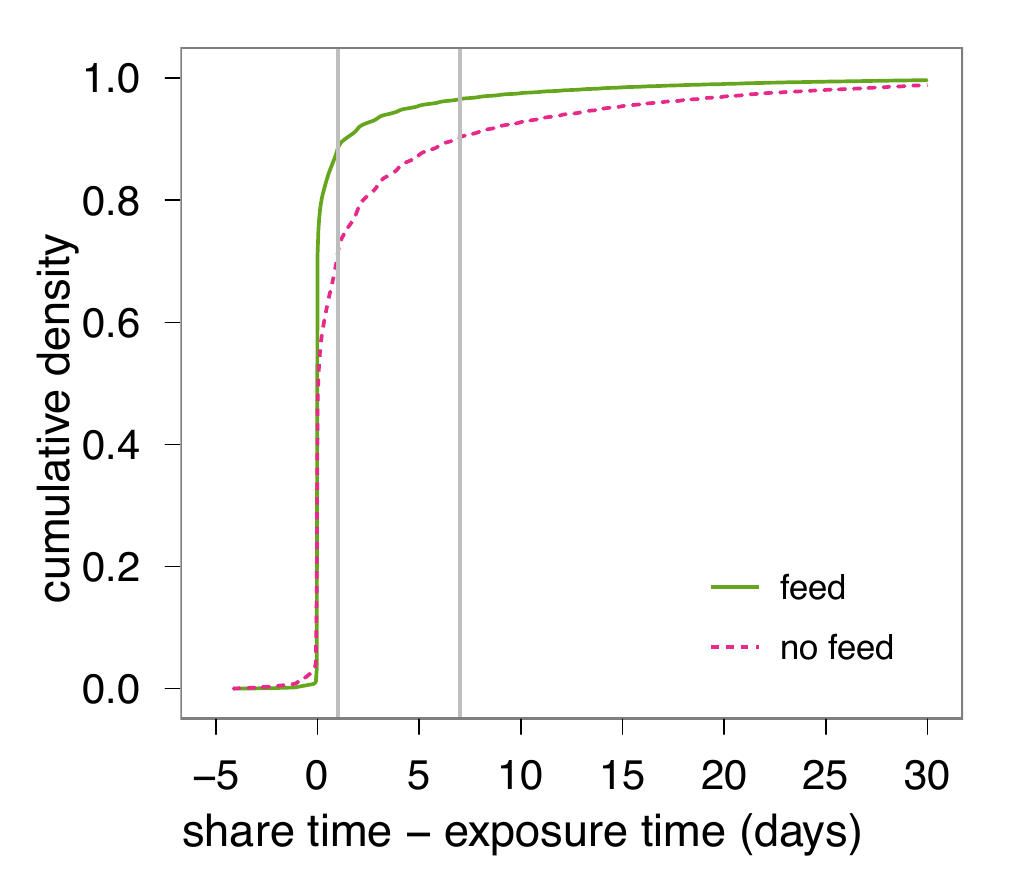}\\
(b) \\
\end{tabular}
\caption{Temporal clustering in sharing the same link as a friend in the {\it feed} and {\it no feed} conditions. 
(a) The difference in sharing time between a subject and their first sharing friend.  (b) The difference between the time at which a subject was first to exposed (or was to be exposed) to the link and the time at which they shared.
Vertical lines indicate one day and one week.}
\label{fig:temporal}
\end{center}
\end{figure}

\subsection{Temporal Clustering}
Contemporaneous behavior among connected individuals is commonly
  used as evidence for social influence processes (e.g.~\cite{Anagnostopoulos08kdd,bakshy2009,bakshy2011wsdm,cha2009,christakis2007,gladwell_00,Gomez2010,gruhl2004information,lerman2010information,onnela2010,wu2011www}).  We find that subjects who share the same link as their friends typically do so within a time that is proximate to their friends' sharing time, even when no exposure occurs on Facebook.  Figure~\ref{fig:temporal} illustrates the cumulative distribution of information lags between the subject and their first sharing friend, among subjects who had shared a URL after their friends.  The top panel shows the latency in sharing times between the subject and their friend for users in the {\it feed} and {\it no feed} condition.  While a larger proportion of users in the {\it feed} condition share a link within the first hour of their friends, the distribution of sharing times is strikingly similar.  The bottom panel shows the differences in time between when subjects shared and when they were (or would have been) first exposed to their friends' sharing behavior on the Facebook feed. The horizontal axis is negative when a subject had shared a link after a friend but had not yet seen that link on the feed.  From this comparison, it is easy to see that users in the {\it feed} condition are most likely to share a link immediately upon exposure, while those who share it without seeing it in their feed will do so over a slightly longer period of time.

To evaluate how exposure on the Facebook feed relates to the speed at which URLs appear to diffuse, we consider URLs that were assigned to both the {\it feed} and {\it no feed} condition. We first match the share time of each URL in the {\it feed} condition with a share time of the URL in the {\it no feed} condition, sampling URLs in proportion to their relative abundances in the data.  From this set of contrasts, we find that the median sharing latency after a friend has already shared the content is 6 hours in the {\it feed} condition, compared to 20 hours when assigned to the {\it no feed} condition (Wilcoxon rank-sum test, $p < 10^{-16}$).  The presence of strong temporal clustering in both experimental conditions illustrates the problem with inferring influence processes from observations of temporally proximate behavior among connected individuals: regardless of access to social signals within a particular online medium, individuals can still acquire and share the same information as their friends, albeit at a slightly later point in time.

\subsection{Effect of Multiple Sharing Friends}

\begin{figure*}[!htb]
\begin{center}
\centering
\subfloat[]{\includegraphics[width=0.32\textwidth]{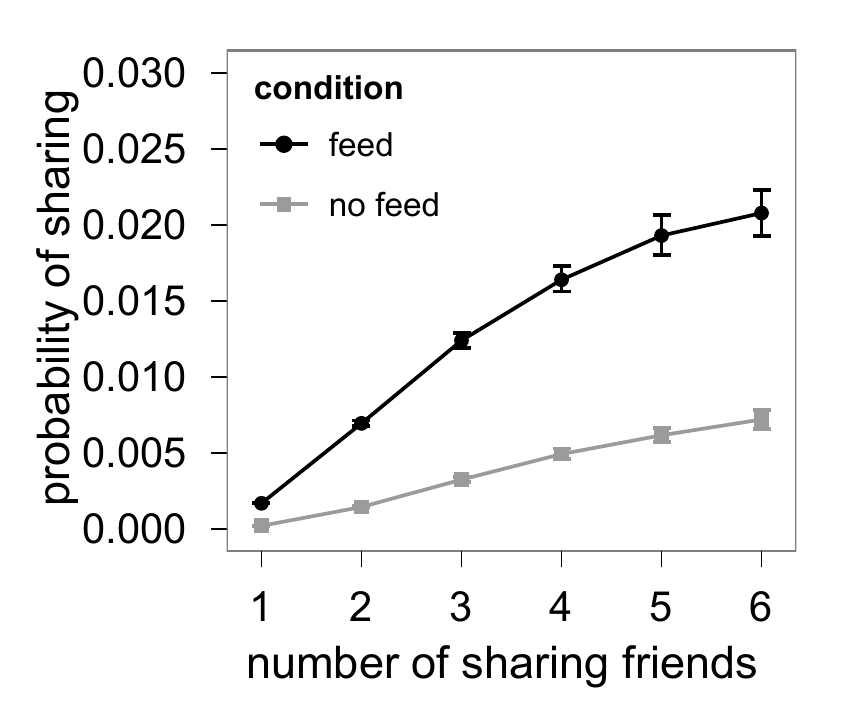}}
\subfloat[]{\includegraphics[width=0.32\textwidth]{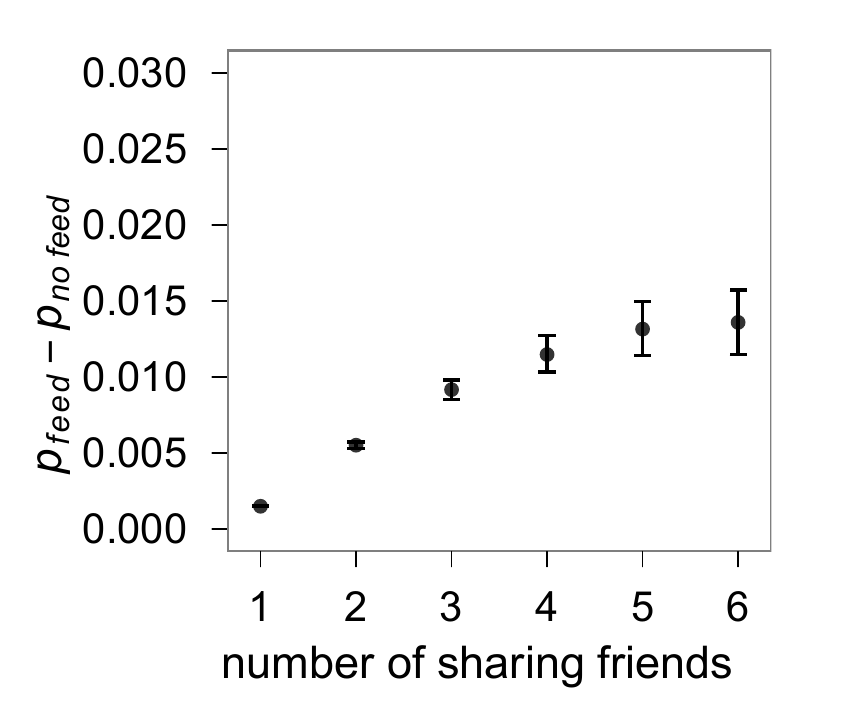}}
\subfloat[]{\includegraphics[width=0.32\textwidth]{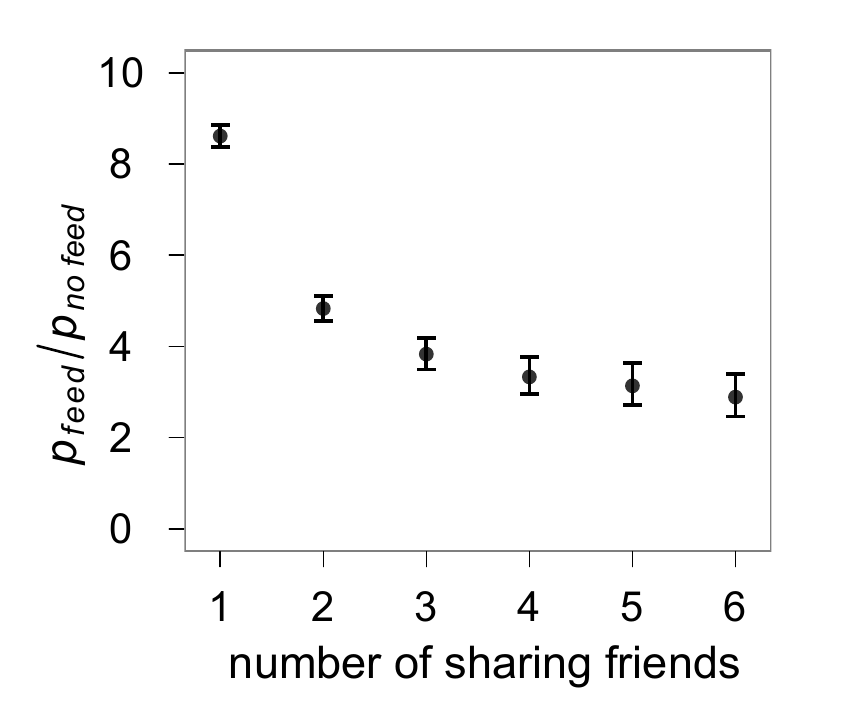}}
\caption{Users with more friends sharing a Web link are themselves more likely to share.
(a) The probability of sharing for subjects that were ({\it feed}) and were not ({\it no feed}) exposed to content increases as a function of the number sharing friends.  
(b) The causal effect of the feed is greater when subjects have more sharing friends
(c) The multiplicative impact of the feed is greatest when few friends are sharing. Error bars represent the 95\% bootstrapped confidence intervals clustered on the URL.
}
\label{fig:numfriends}
\end{center}
\end{figure*}
Classic models of social and biological contagion (e.g.~\cite{granovetter1978,newman2002epidemic}) predict that the likelihood of ``infection'' increases with the number of infected contacts. Observational studies of online contagion~\cite{Anagnostopoulos08kdd,bakshy2009,cha2009,leskovec06viral} not only find evidence of temporal clustering, but also observe a similar relationship between the likelihood of contagion and the number of infected contacts.  However, it is important to note that this correlation can have multiple causes that are unrelated to social influence processes.  For example, if a website is popular among friends, then a particularly interesting page is more likely to be shared by a users' friends independent of one another.  The positive relationship between the number of sharing friends and likelihood of sharing may therefore simply reflect heterogeneity in the ``interestingness'' of the content, which is clustered along the network: the more popular a page is for a group of friends, the more likely it is that one would observe multiple friends sharing it.

We first show that, consistent with prior observational studies, the probability of sharing a link in the {\it feed} condition increases with the number of contacts who have already shared the link (solid line, Figure~\ref{fig:numfriends}a).
But the presence of a similar relationship in the {\it no feed}  condition (grey line, Figure~\ref{fig:numfriends}a) shows that an individual is more likely to exhibit the sharing behavior when multiple friends share, even if she does not necessarily observe her friends' behavior.  Therefore, when using observational data, the na\"{\i}ve conditional probability (which is equivalent to the probability of sharing in the {\it feed} condition) does not directly give the probability increase due to influence via multiple sharing friends. Rather, such an estimate reflects a mixture of internal influence effects and external correlation.

Our experiment allows us to directly measure the effect of the feed relative to external factors, computed as either the difference or ratio between the probability of sharing in the {\it feed} and {\it no feed} conditions (Figure~\ref{fig:numfriends}bc). 
While the difference in sharing likelihood grows with the number of sharing friends, the relative risk ratio falls.
This contrast suggests that social information in the feed is most likely to influence a user to share a link that many of her friends have shared, but the relative impact of that influence is highest for content that few friends are sharing.  The decreasing relative effect is consistent with the hypothesis that having multiple sharing friends is associated with greater redundancy in information exposure, which may either be caused by homophily in visitation and sharing tendencies, or external influence.

\section{Tie Strength and Influence}\label{sec:tie_strength}
\begin{figure}[th]
\begin{center}
\includegraphics[width=0.45\textwidth]{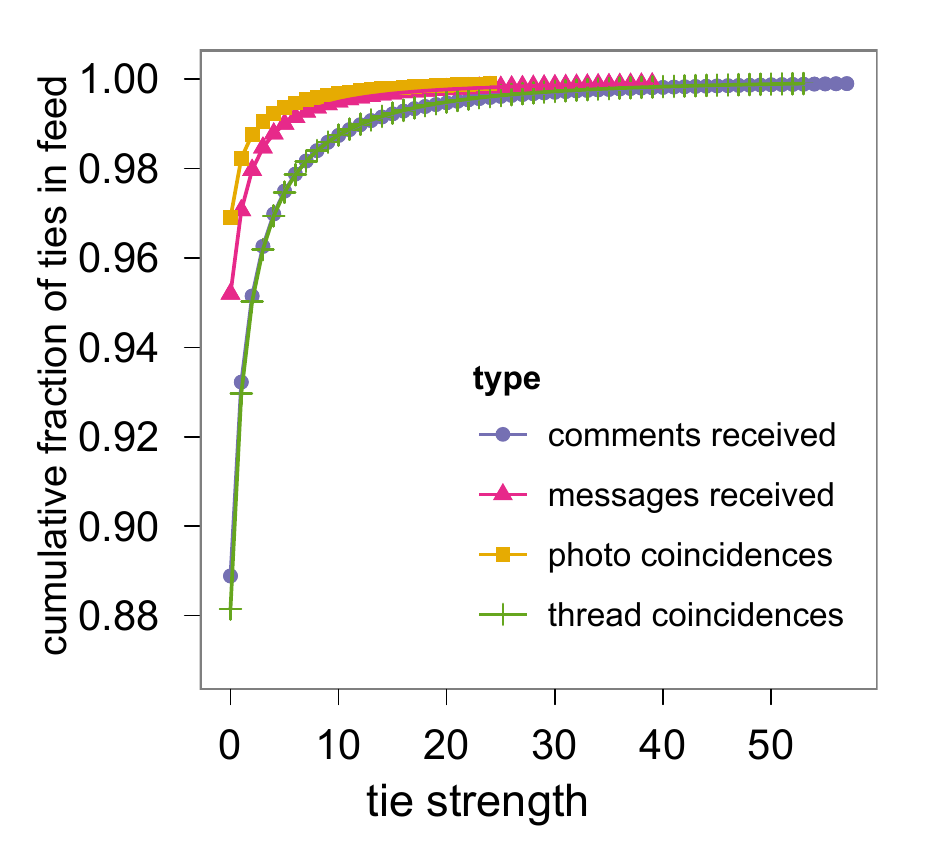}
\caption{Tie strength distribution among friends displayed in subjects' feeds using the four measurements.  Points are plotted up to the $99.9^{th}$ percentile.  Note that the vertical axis is collapsed.}
\label{fig:ts_dist}
\end{center}
\end{figure}

Next, we examine the relationship between tie strength, influence, and
information diversity by combining the experimental data with users'
online and offline interactions. Following arguments originally
proposed by Mark Granovetter's seminal 1973 paper, {\it The Strength
  of Weak Ties}~\cite{granovetter1973}, empirical work linking tie
strength and diffusion often utilize the number of mutual
contacts as proxies of interaction frequency. Rather than
  using the number of mutual contacts, which can be large for
  pairs of individuals who no longer communicate (e.g. former
  classmates), we directly measure the strength of tie between a subject and her friend in terms of four types of interactions: (i) the frequency of private online communication between the two users in the form of Facebook messages\footnote{We quantify message and comment interactions as the number of communication events the subject received from their friend. The number of messages and comments sent, and the geometric mean of communications sent and received, yielded qualitatively similar results, so we plot only the single directed measurement for the sake of clarity.}; (ii) the frequency of public online interaction in the form of comments left by one user on another user's posts; (iii) the number of real-world coincidences captured on Facebook in terms of both users being labeled by users as appearing in the same photograph; and (iv) the number of online coincidences in terms of both users responding to the same Facebook post with a comment.  Frequencies are computed using data from the three months directly prior to the experiment.  The distribution of tie strengths among subjects and their sharing friends can be seen in Figure~\ref{fig:ts_dist}.

\begin{figure*}[!htb]
\centering
\subfloat[\emph{}]{
  \includegraphics[width=0.24\textwidth]{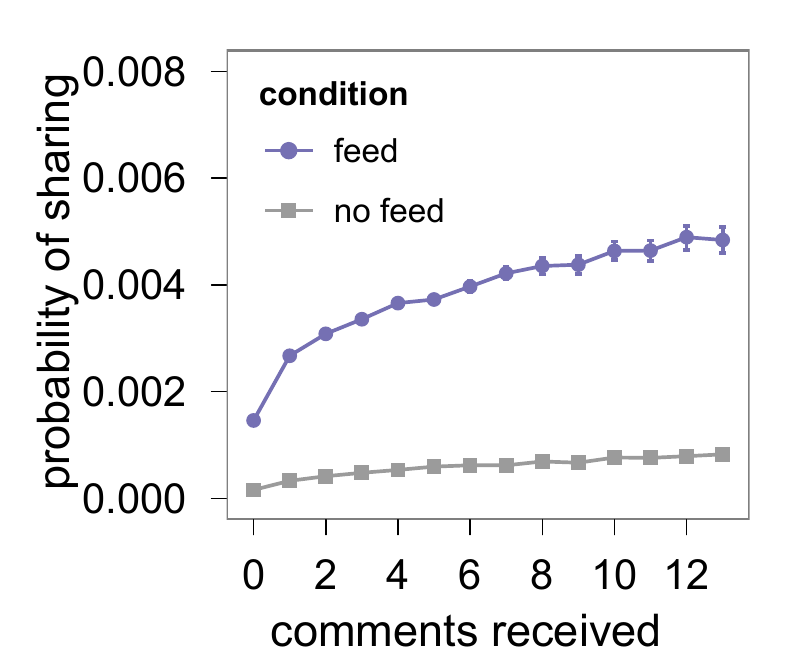}
  \includegraphics[width=0.24\textwidth]{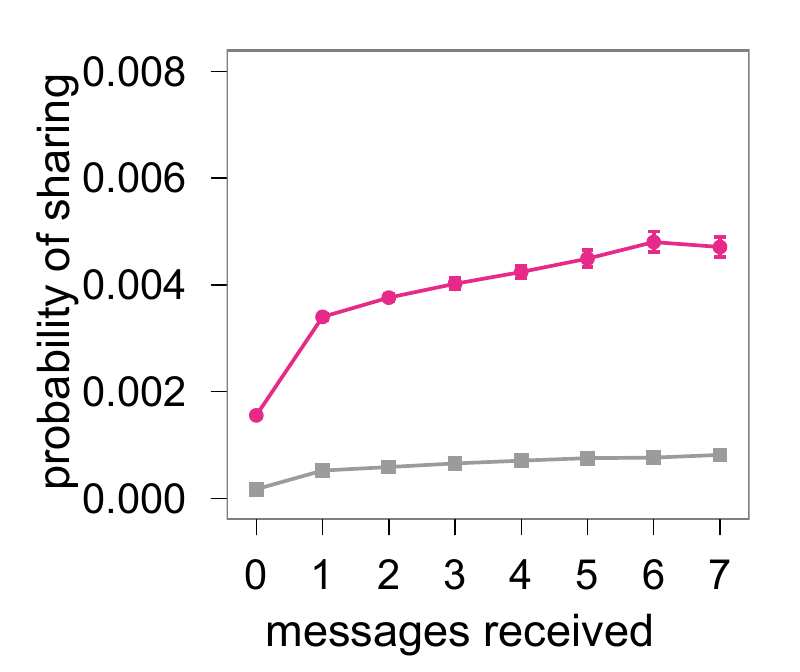}
  \includegraphics[width=0.24\textwidth]{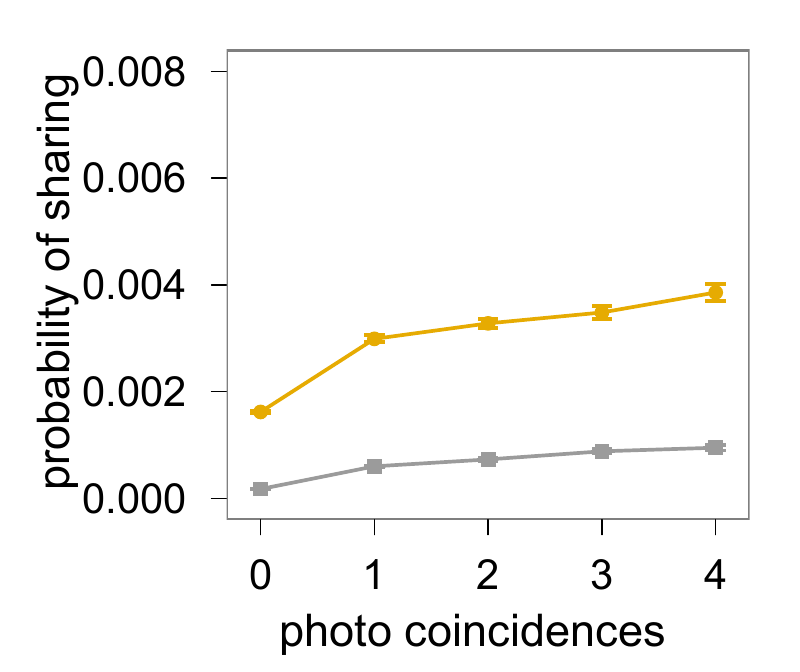}
  \includegraphics[width=0.24\textwidth]{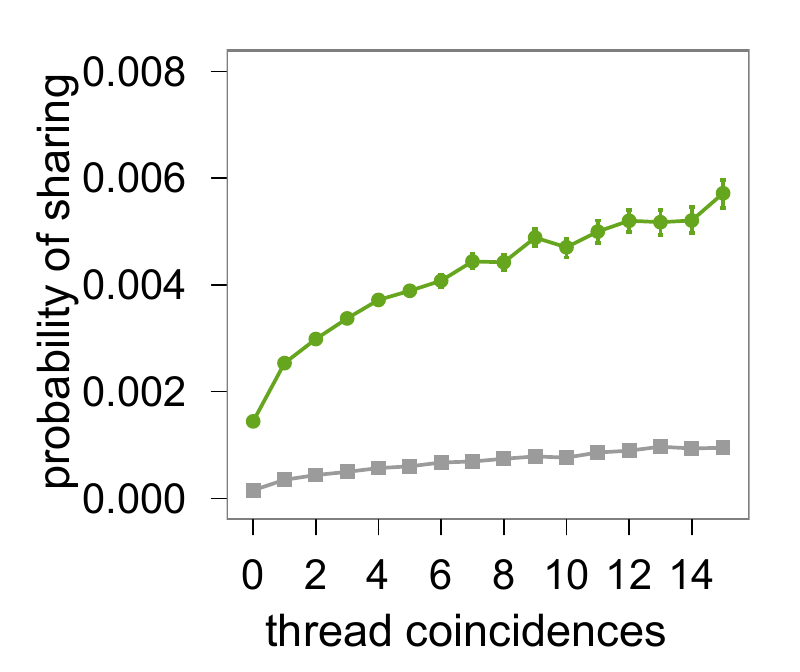}
   \label{fig:ts_pr}
}\\
\subfloat[\emph{}]{
\includegraphics[width=0.24\textwidth]{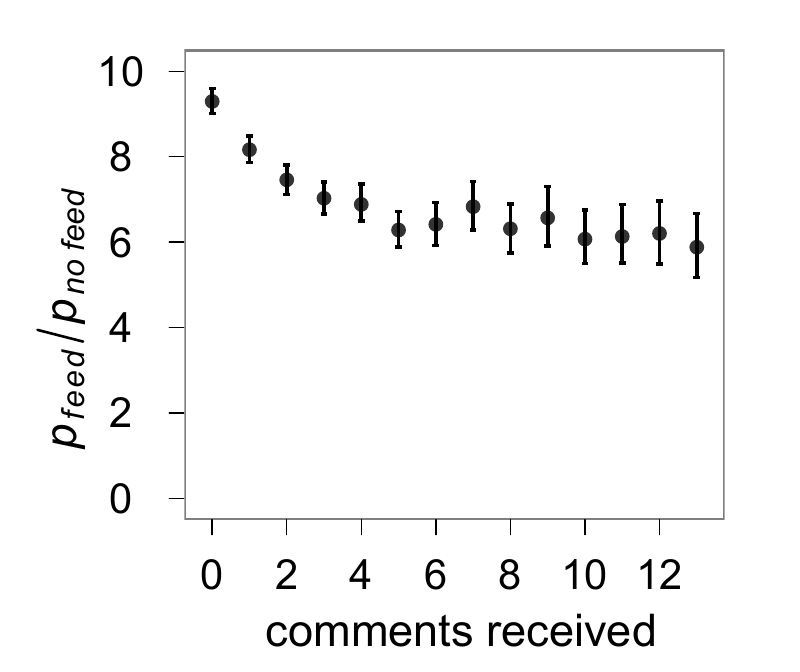}
\includegraphics[width=0.24\textwidth]{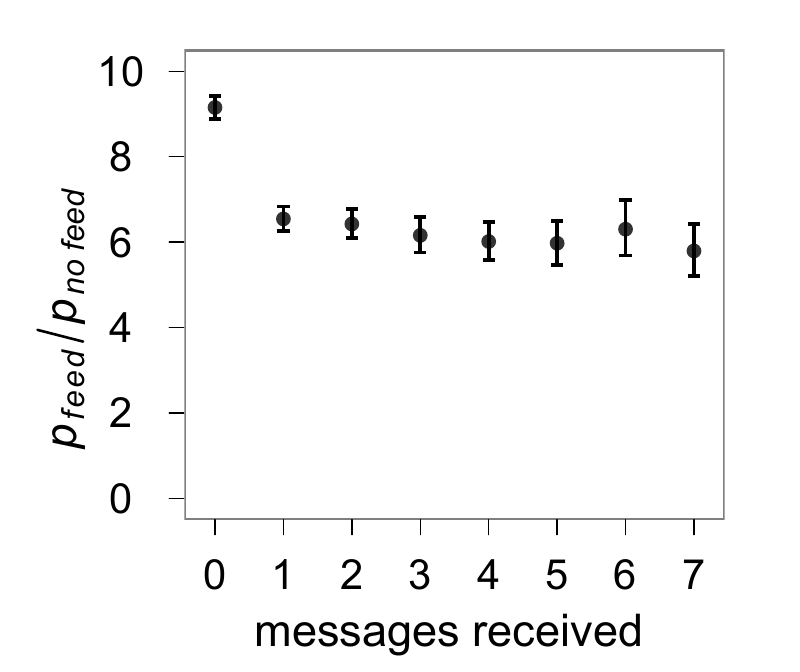}
\includegraphics[width=0.24\textwidth]{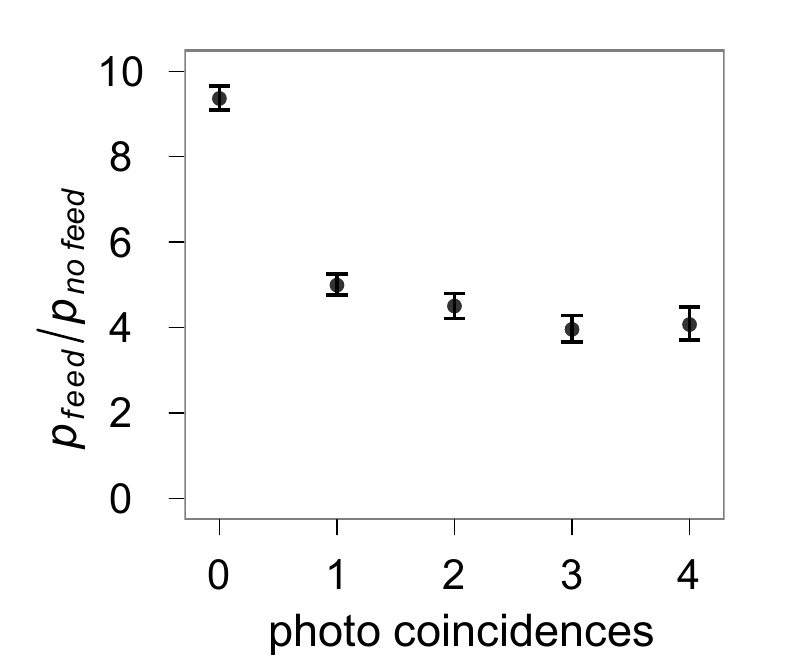}
\includegraphics[width=0.24\textwidth]{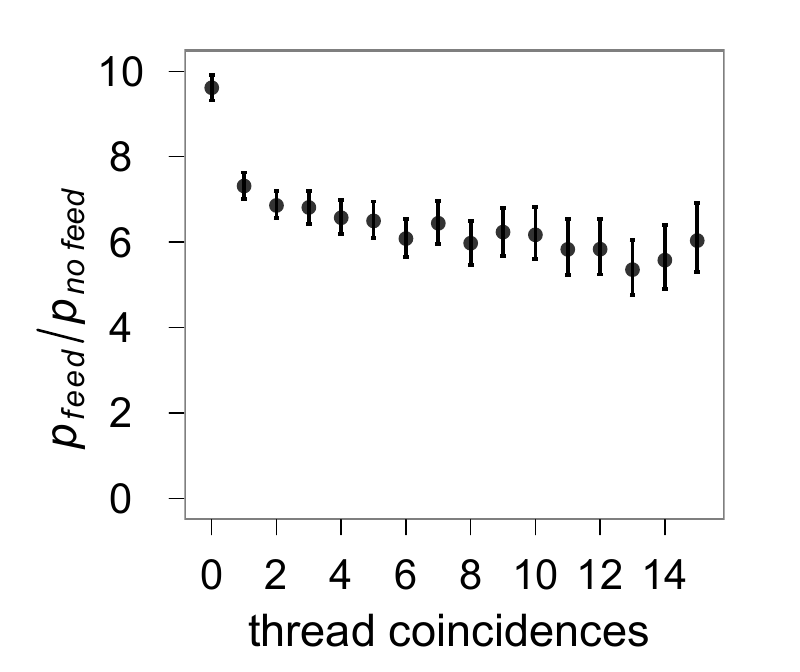}
 \label{fig:ts_rr}
}
\caption{Strong ties are more influential, and weak ties expose friends to information they would not have otherwise shared.
(a) The increasing relationship between tie strength and the probability of sharing a link that a friend shared in the {\it feed} and {\it no feed} conditions.  (b) The multiplicative effect of feed diminishes with tie strength, suggesting that exposure through strong ties may be redundant with external exposure, while weak ties carry information one might otherwise not have been exposed to.
}
\label{fig:tie_strength_effect}
\end{figure*}

\subsection{Effect of Tie Strength}
We measure how the difference in the likelihood of sharing a URL in the {\it feed} versus {\it no feed} conditions varies according to tie strength. To simplify our estimate of the effect of tie strength, we restrict our analysis to subjects with exactly one friend who had previously shared the link.
In both conditions, a subject is more likely to share a link when her sharing friend is a strong tie (Figure~\ref{fig:ts_pr}).
For example, subjects who were exposed to a link shared by a friend from whom the subject received three comments are 2.83 times more likely to share than subjects exposed to a link shared by a friend from whom they received no comments.
For those who were not exposed, the same comparison shows that subjects are 3.84 times more likely to share a link that was previously shared by the stronger tie.  The larger effect in the {\it no feed} condition suggests that tie strength is a stronger predictor of externally correlated activity than it is for influence on feed.  From Figure~\ref{fig:ts_pr}, it is also clear that individuals are more likely to be influenced by their stronger ties via the feed to share content that they would not have otherwise spread.

Furthermore, our results extend Granovetter's hypothesis that weak ties disseminate novel information into the context of media contagion.  Figure~\ref{fig:ts_rr} shows that the risk ratio of sharing between the {\it feed} and {\it no feed} conditions is highest for content shared by weak ties.
This suggests that weak ties consume and transmit information that one is unlikely to be exposed to otherwise, thereby increasing the diversity of information propagated within the network.

\subsection{Collective Impact of Ties}
Strong ties may be individually more influential, but how much diffusion occurs  in aggregate through these ties depends on the underlying distribution of tie strength (i.e. Figure~\ref{fig:ts_dist}).
Using the experimental data, we can estimate the amount of contagion
on the feed generated by strong and weak ties. The causal
effect of exposure to information shared by friends with tie strength
$k$ is given by the average treatment effect on the treated:
\[\textrm{\it ATET(k)} = p(k, \textrm{\it feed}) -
p(k, \textrm{\it no feed})
\]

To determine the collective impact of ties of strength $k$, we multiply this quantity by the fraction of links displayed in all users' feeds posted by friends of tie strength $k$, denoted by $f(k)$.  In order to compare the impact of weak and strong ties, we must set a cutoff value for the minimum amount of interaction required between two individuals in order to consider that tie strong. Setting the cutoff at $k=1$ (a single interaction) provides the most generous classification of strong ties while preserving some meaningful distinction between strong and weak ties, thereby giving the most influence credit to strong ties.

\begin{figure}[tbh]
\begin{center}
\includegraphics[width=.42\textwidth]{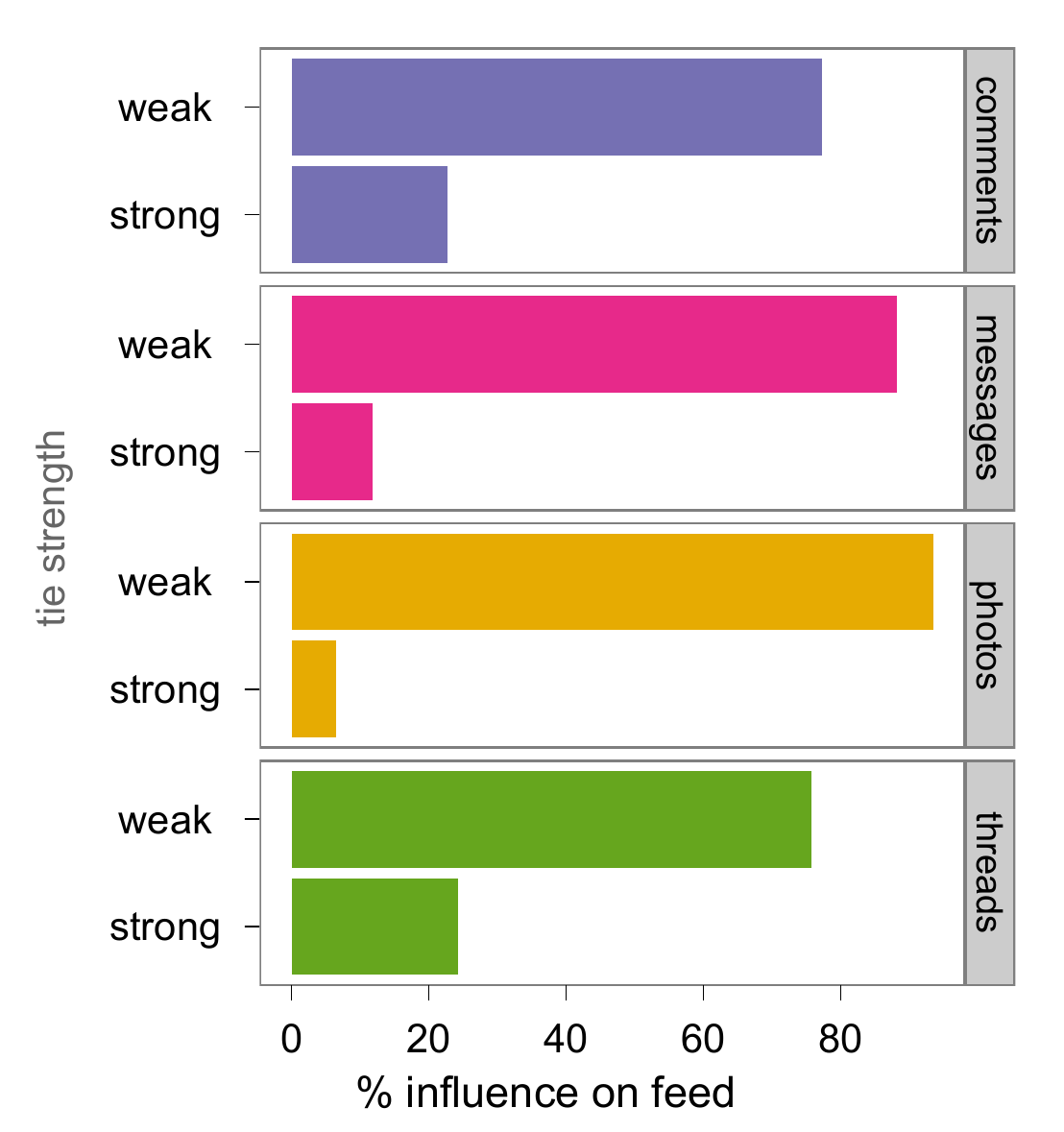}
\caption{Weak ties are collectively more influential than strong ties.
Panels show the percentage of information spread by strong and weak ties for all four measurements of tie strength.
Although the probability of influence is significantly higher for those that interact frequently, most contagion occurs along weak ties, which are more abundant.
}
\label{fig:influence_volume}
\end{center}
\end{figure}

Under this categorization of strong and weak ties, the estimated total
fraction of sharing events that can be attributed to weak and strong
ties is 
the average treatment effect on the treated weighted by the proportion
of URL exposures from each tie type:
\begin{align*}
  &T_{\textrm{\it \footnotesize{weak}}} = \textrm{\it ATET} (0)*f(0) \\
  &T_{\textrm{\it \footnotesize{strong}}}  = \sum_{i=1}^{N}{\textrm{\it
        ATET} (i)*f(i)}
\end{align*}
We illustrate this comparison in Figure~\ref{fig:influence_volume}, and show that by a wide margin,  the majority of influence is generated by weak ties\footnote{Note that for the purposes of this study, it is not necessary to model the effect of tie strength for users with multiple sharing friends, since stories of this kind only constitute 4.2\% of links in the newsfeed, and their inclusion would not dramatically alter the balance of aggregate influence by tie strength.}.  Although we have shown that strong ties are individually more influential, the effect of strong ties is not large enough to match the sheer abundance of weak ties.

\section{Discussion}\label{sec:discussion}
Social networks may influence an individual's behavior, but they also reflect the individual's own activities, interests, and opinions.
These commonalities make it nearly impossible to determine from observational data whether any particular interaction, mode of communication, or social environment is responsible for the apparent spread of a behavior through a network.
In the context of our study, there are three possible mechanisms that may explain diffusion-like phenomena:
(1) An individual shares a link on Facebook, and exposure to this information on the feed causes a friend to re-share that same link.
(2) Friends visit the same web page and share a link to that web page on Facebook, independently of one another.
(3) An individual shares a link within and external to Facebook, and exposure to the externally shared information causes a friend to share the link on Facebook.
Our experiment determines the causal effect of the feed on the spread of sharing behaviors by comparing the likelihood of sharing under the {\it feed} condition (possible causes 1-3) with the likelihood under the {\it no feed} condition (possible causes 2-3).

Our experiment generalizes Mark Granovetter's predictions about the strength of weak ties~\cite{granovetter1973} to the spread of everyday information. Weak ties are argued to have access to more diverse information because they are expected to have fewer mutual contacts;
each individual has access to information that the other does not.  For information that is almost exclusively embedded within few individuals, like job openings or future strategic plans, weak ties play a necessarily role in facilitating information flow.  This reasoning, however, does not necessarily apply to the spread of widely available information, and the relationship between tie strength and information access is not immediately obvious.
Our experiment sheds light on how tie strength relates to information access within a broader context, and suggests that weak ties, defined directly in terms of interaction propensities, diffuse novel information that would not have otherwise spread.

Although weak ties can serve a critical bridging function~\cite{granovetter1973,onnela2007}, the influence that weak ties exert has never before been measured empirically at a systemic level. We find that the majority of influence results from exposure to individual weak ties, which indicates that most information diffusion on Facebook is driven by simple contagion.
This stands in contrast to prior studies of influence on the adoption of products, behaviors or opinions, which center around the effect of having multiple or densely connected contacts who have adopted~\cite{aral2011creating,backstrom06,centola2007,centola2010}. Our results suggest that in large online environments, the low cost of disseminating information fosters diffusion dynamics that are different from situations where adoption is subject to positive externalities or carries a high cost.

Because we are unable to observe interactions that occur outside of Facebook, a limitation of our study is that we can only fully identify causal effects within the site.  Correlated sharing in the {\it no feed} condition may occur because friends independently visit and share the same page as one another, or because one user is influenced to share via an external communication channel.  Although we are not able to directly evaluate the relative contribution of these two potential causes, our results allow us to obtain a bound on the effect on sharing behavior within the site.  The probability of sharing in the {\it no feed} condition, which is a combination of similarity and external influence, is an upper bound on how much sharing occurs because of homophily-related effects.  Likewise, the difference in the probability of sharing within the {\it feed} and {\it no feed} condition gives a lower bound on how much on-site sharing is due to interpersonal influence along any communication medium.

The mass adoption of online social networking systems has the potential to dramatically alter an individual's exposure to new information. By applying an experimental approach to measuring diffusion outcomes within one of the largest human communication networks, we are able to rigorously quantify the effect of social networks on information spread.
The present work sheds light on aggregate trends over a large population; future studies may investigate how properties of the individual, such as age, gender, and nationality, or features of content, such as popularity and breadth of appeal, relate to the influence and its confounds.

\section{Acknowledgments}
We would like to thank Michael D. Cohen, Dean Eckles, Emily Falk, James Fowler, and Brian Karrer for their discussions and feedback on this work. This work was supported in part by NSF IIS-0746646.

\bibliographystyle{abbrv}

\end{document}